\documentclass[twocolumn]{aastex63}
\hypersetup{linkcolor=blue,urlcolor=blue}
\accepted{April 11, 2023}

\shorttitle{JWST Reveals a Filamentary Structure around a $z=6.61$ Quasar}
\shortauthors{Wang et al.}
\graphicspath{{./}{figures/}}

\begin{document}

\title{A SPectroscopic survey of biased halos In the Reionization Era (ASPIRE): JWST Reveals a Filamentary Structure around a $z=6.61$ Quasar}

\correspondingauthor{Feige Wang}
\email{feigewang@email.arizona.edu}

\author[0000-0002-7633-431X]{Feige Wang}
\affiliation{Steward Observatory, University of Arizona, 933 N Cherry Avenue, Tucson, AZ 85721, USA}

\author[0000-0001-5287-4242]{Jinyi Yang}
\thanks{Strittmatter Fellow}
\affiliation{Steward Observatory, University of Arizona, 933 N Cherry Avenue, Tucson, AZ 85721, USA}

\author[0000-0002-7054-4332]{Joseph F.\ Hennawi}
\affiliation{Department of Physics, University of California, Santa Barbara, CA 93106-9530, USA}
\affiliation{Leiden Observatory, Leiden University, Niels Bohrweg 2, NL-2333 CA Leiden, Netherlands}

\author[0000-0003-3310-0131]{Xiaohui Fan}
\affiliation{Steward Observatory, University of Arizona, 933 N Cherry Avenue, Tucson, AZ 85721, USA}

\author[0000-0002-4622-6617]{Fengwu Sun}
\affiliation{Steward Observatory, University of Arizona, 933 N Cherry Avenue, Tucson, AZ 85721, USA}

\author[0000-0002-6184-9097]{Jaclyn B. Champagne}
\affiliation{Steward Observatory, University of Arizona, 933 N Cherry Avenue, Tucson, AZ 85721, USA}

\author[0000-0002-6748-2900]{Tiago Costa}
\affiliation{Max-Planck-Institut f\"ur Astrophysik, Karl-Schwarzschild-Stra{\ss}e 1, D-85748 Garching b. M\"unchen, Germany}

\author[0000-0003-4750-0187]{Melanie Habouzit}
\affiliation{Zentrum f\"ur Astronomie der Universit\"at Heidelberg, ITA, Albert-Ueberle-Str. 2, D-69120 Heidelberg, Germany}
\affiliation{Max-Planck-Institut f\"ur Astronomie, K\"onigstuhl 17, D-69117 Heidelberg, Germany}

\author[0000-0003-4564-2771]{Ryan Endsley}
\affiliation{Department of Astronomy, The University of Texas at Austin, Austin, TX 78712, USA}

\author[0000-0001-5951-459X]{Zihao Li}
\affiliation{Department of Astronomy, Tsinghua University, Beijing 100084, China}

\author[0000-0001-6052-4234]{Xiaojing Lin}
\affiliation{Department of Astronomy, Tsinghua University, Beijing 100084, China}

\author[0000-0001-5492-4522]{Romain A. Meyer}
\affiliation{Max Planck Institut f\"ur Astronomie, K\"onigstuhl 17, D-69117, Heidelberg, Germany}

\author[0000-0002-4544-8242]{Jan--Torge Schindler}
\affiliation{Leiden Observatory, Leiden University, Niels Bohrweg 2, NL-2333 CA Leiden, Netherlands}

\author[0000-0003-0111-8249]{Yunjing Wu}
\affiliation{Department of Astronomy, Tsinghua University, Beijing 100084, China}
\affiliation{Steward Observatory, University of Arizona, 933 N Cherry Avenue, Tucson, AZ 85721, USA}

\author[0000-0002-2931-7824]{Eduardo Ba\~nados}
\affil{Max Planck Institut f\"ur Astronomie, K\"onigstuhl 17, D-69117, Heidelberg, Germany}

\author[0000-0002-3026-0562]{Aaron J. Barth}
\affil{Department of Physics and Astronomy, 4129 Frederick Reines Hall, University of California, Irvine, CA, 92697-4575, USA}

\author[0000-0002-7080-2864]{Aklant K. Bhowmick}
\affiliation{Department of Physics, University of Florida, Gainesville, FL, 32611, USA}

\author{Rebekka Bieri}
\affil{Center for Space and Habitability, University of Bern, Gesellschaftsstrasse 6 (G6), Bern, Switzerland}

\author[0000-0002-2183-1087]{Laura Blecha}
\affiliation{Department of Physics, University of Florida, Gainesville, FL, 32611, USA}

\author[0000-0001-8582-7012]{Sarah Bosman}
\affiliation{Max Planck Institut f\"ur Astronomie, K\"onigstuhl 17, D-69117, Heidelberg, Germany}

\author[0000-0001-8467-6478]{Zheng Cai}
\affiliation{Department of Astronomy, Tsinghua University, Beijing 100084, China}

\author{Luis Colina}
\affil{Centro de Astrobiología (CAB), CSIC-INTA, Ctra. de Ajalvir km 4, Torrejón de Ardoz, E-28850, Madrid, Spain}
\affil{International Associate, Cosmic Dawn Center (DAWN)}

\author[0000-0002-7898-7664]{Thomas Connor}
\affiliation{Center for Astrophysics $\vert$\ Harvard\ \&\ Smithsonian, 60 Garden St., Cambridge, MA 02138, USA}
\affiliation{Jet Propulsion Laboratory, California Institute of Technology, 4800 Oak Grove Drive, Pasadena, CA 91109, USA}

\author[0000-0003-0821-3644]{Frederick B.\ Davies}
\affiliation{Max Planck Institut f\"ur Astronomie, K\"onigstuhl 17, D-69117, Heidelberg, Germany}

\author[0000-0002-2662-8803]{Roberto Decarli}
\affil{INAF--Osservatorio di Astrofisica e Scienza dello Spazio, via Gobetti 93/3, I-40129, Bologna, Italy}

\author[0000-0003-3242-7052]{Gisella De Rosa}
\affiliation{Space Telescope Science Institute, 3700 San Martin Dr, Baltimore, MD 21210}

\author[0000-0002-0174-3362]{Alyssa B.\ Drake}
\affiliation{Centre for Astrophysics Research, Department of Physics, Astronomy and Mathematics, University of Hertfordshire, Hatfield AL10 9AB, UK}

\author[0000-0003-1344-9475]{Eiichi Egami}
\affiliation{Steward Observatory, University of Arizona, 933 N Cherry Avenue, Tucson, AZ 85721, USA}

\author[0000-0003-2895-6218]{Anna-Christina Eilers}\thanks{Pappalardo Fellow}
\affiliation{MIT Kavli Institute for Astrophysics and Space Research, 77 Massachusetts Avenue, Cambridge, 02139, Massachusetts, USA}

\author[0000-0003-0850-7749]{Analis E. Evans}
\affiliation{Department of Physics, University of Florida, Gainesville, FL, 32611, USA}

\author[0000-0002-6822-2254]{Emanuele Paolo Farina}
\affiliation{Gemini Observatory, NSF's NOIRLab, 670 N A'ohoku Place, Hilo, Hawai'i 96720, USA}

\author[0000-0003-3633-5403]{Zoltan Haiman}
\affil{Department of Astronomy, Columbia University, New York, NY 10027, USA}
\affil{Department of Physics, Columbia University, New York, NY 10027, USA}

\author[0000-0003-4176-6486]{Linhua Jiang}
\affiliation{Department of Astronomy, School of Physics, Peking University, Beijing 100871, China}
\affiliation{Kavli Institute for Astronomy and Astrophysics, Peking University, Beijing 100871, China}

\author[0000-0002-5768-738X]{Xiangyu Jin}
\affiliation{Steward Observatory, University of Arizona, 933 N Cherry Avenue, Tucson, AZ 85721, USA}

\author[0000-0003-1470-5901]{Hyunsung D. Jun}
\affil{SNU Astronomy Research Center, Seoul National University, 1 Gwanak-ro, Gwanak-gu, Seoul 08826, Republic of Korea}

\author[0000-0001-6874-1321]{Koki Kakiichi}
\affiliation{Department of Physics, Broida Hall, University of California, Santa Barbara, CA 93106-9530, USA}

\author[0000-0002-7220-397X]{Yana Khusanova}
\affil{Max Planck Institut f\"ur Astronomie, K\"onigstuhl 17, D-69117, Heidelberg, Germany}

\author[0000-0001-5829-4716]{Girish Kulkarni}
\affiliation{Tata Institute of Fundamental Research, Homi Bhabha Road, Mumbai 400005, India }

\author[0000-0001-6251-649X]{Mingyu Li}
\affil{Department of Astronomy, Tsinghua University, Beijing 100084, China}

\author[0000-0003-3762-7344]{Weizhe Liu}
\affiliation{Steward Observatory, University of Arizona, 933 N Cherry Avenue, Tucson, AZ 85721, USA}

\author{Federica Loiacono}
\affil{INAF--Osservatorio di Astrofisica e Scienza dello Spazio, via Gobetti 93/3, I-40129, Bologna, Italy}

\author{Alessandro Lupi}
\affil{Dipartimento di Fisica ``G. Occhialini'', Universit\`a degli Studi di Milano-Bicocca, Piazza della Scienza 3, I-20126 Milano, Italy}

\author[0000-0002-5941-5214]{Chiara Mazzucchelli}
\affiliation{Instituto de Estudios Astrof\'{\i}sicos, Facultad de Ingenier\'{\i}a y Ciencias, Universidad Diego Portales, Avenida Ejercito Libertador 441, Santiago, Chile.}

\author[0000-0003-2984-6803]{Masafusa Onoue}
\affiliation{Kavli Institute for Astronomy and Astrophysics, Peking University, Beijing 100871, China}
\affiliation{Kavli Institute for the Physics and Mathematics of the Universe (Kavli IPMU, WPI), The University of Tokyo, Chiba 277-8583, Japan}

\author[0000-0003-4924-5941]{Maria A. Pudoka}
\affiliation{Steward Observatory, University of Arizona, 933 N Cherry Avenue, Tucson, AZ 85721, USA}

\author[0000-0003-2349-9310]{Sof\'ia Rojas-Ruiz}
\altaffiliation{Fellow of the International Max Planck Research School for Astronomy and Cosmic Physics at the University of Heidelberg (IMPRS--HD)}
\affiliation{Max-Planck-Institut f\"{u}r Astronomie, K\"{o}nigstuhl 17, D-69117, Heidelberg, Germany}

\author[0000-0003-1659-7035]{Yue Shen}
\affiliation{Department of Astronomy, University of Illinois at Urbana-Champaign, Urbana, IL 61801, USA}
\affiliation{National Center for Supercomputing Applications, University of Illinois at Urbana-Champaign, Urbana, IL 61801, USA}

\author[0000-0002-0106-7755]{Michael A. Strauss}
\affiliation{Department of Astrophysical Sciences, Princeton University, Princeton, NJ 08544 USA}

\author[0000-0003-0747-1780]{Wei Leong Tee}
\affiliation{Steward Observatory, University of Arizona, 933 N Cherry Avenue, Tucson, AZ 85721, USA}

\author[0000-0002-3683-7297]{Benny Trakhtenbrot}
\affiliation{School of Physics and Astronomy, Tel Aviv University, Tel Aviv 69978, Israel}

\author[0000-0002-6849-5375]{Maxime Trebitsch}
\affil{Kapteyn Astronomical Institute, University of Groningen, P.O. Box 800, 9700 AV Groningen, The Netherlands}

\author[0000-0001-9024-8322]{Bram Venemans}
\affiliation{Leiden Observatory, Leiden University, Niels Bohrweg 2, NL-2333 CA Leiden, Netherlands}

\author{Marta Volonteri}
\affil{Institut d'Astrophysique de Paris, Sorbonne Universit\'e, CNRS, UMR 7095, 98 bis bd Arago, 75014 Paris, France}

\author[0000-0003-4793-7880]{Fabian Walter}
\affiliation{Max Planck Institut f\"ur Astronomie, K\"onigstuhl 17, D-69117, Heidelberg, Germany}

\author[0000-0002-0125-6679]{Zhang-Liang Xie}
\affiliation{Max Planck Institut für Astronomie, Königstuhl 17, D-69117, Heidelberg, Germany}

\author[0000-0002-5367-8021]{Minghao Yue}
\affiliation{MIT Kavli Institute for Astrophysics and Space Research, 77 Massachusetts Ave., Cambridge, MA 02139, USA}
\affiliation{Steward Observatory, University of Arizona, 933 N Cherry Avenue, Tucson, AZ 85721, USA}

\author[0000-0002-4321-3538]{Haowen Zhang}
\affiliation{Steward Observatory, University of Arizona, 933 N Cherry Avenue, Tucson, AZ 85721, USA}

\author[0000-0002-0123-9246]{Huanian Zhang} 
\affil{Department of Astronomy, Huazhong University of Science and Technology, Wuhan, 430074, China}

\author[0000-0002-3983-6484]{Siwei Zou}
\affiliation{Department of Astronomy, Tsinghua University, Beijing 100084, China}

\begin{abstract}
We present the first results from the JWST ASPIRE program (A SPectroscopic survey of biased halos In the Reionization Era). This program represents an imaging and spectroscopic survey of 25 reionization-era quasars and their environments by utilizing the unprecedented capabilities of NIRCam Wide Field Slitless Spectroscopy (WFSS) mode. ASPIRE will deliver the largest ($\sim280~{\rm arcmin}^2$) galaxy redshift survey at 3--4 $\mu$m among JWST Cycle-1 programs and provide extensive legacy values for studying the formation of the earliest supermassive black holes (SMBHs), the assembly of galaxies, early metal enrichment, and cosmic reionization. In this first ASPIRE paper, we report the discovery of a filamentary structure traced by the luminous quasar J0305--3150 and ten [\ion{O}{3}] emitters at $z=6.6$. This structure has a 3D galaxy overdensity of $\delta_{\rm gal}=12.6$ over 637 cMpc$^3$, one of the most overdense structures known in the early universe, and could eventually evolve into a massive galaxy cluster. Together with existing VLT/MUSE and ALMA observations of this field, our JWST observations reveal that J0305–3150 traces a complex environment where both UV-bright and dusty galaxies are present, and indicate that the early evolution of galaxies around the quasar is not simultaneous. In addition, we discovered 31 [\ion{O}{3}] emitters in this field at other redshifts, $5.3<z<6.7$, with half of them situated at $z\sim5.4$ and $z\sim6.2$. This indicates that star-forming galaxies, such as [\ion{O}{3}] emitters, are generally clustered at high redshifts. These discoveries demonstrate the unparalleled redshift survey capabilities of NIRCam WFSS and the potential of the full ASPIRE survey dataset. 
\end{abstract}

\keywords{Early universe (435) --- Galaxies(573) --- Protoclusters (1297) --- Redshift surveys (1378) --- Supermassive black holes (1663)}

\section{Introduction}
Quasars, powered by accreting supermassive black holes (SMBHs) with masses of $10^8-10^{10}~M_\odot$, have been observed up to $z=7.6$ \citep{Banados18a, Yang20a, Wang21a}, deep into the epoch of reionization (EoR). 
How these quasars formed within the first billion years after the Big Bang is one of the most important open questions in astrophysics. Cosmological simulations suggest that billion-solar-mass SMBHs in the EoR formed in massive dark matter halos \citep[e.g.,][]{Springel05,DiMatteo05} and grew through cold flow accretion \citep[e.g.,][]{DiMatteo12} and/or merging with other gas-rich halos \citep[e.g.,][]{Li07}. As a consequence, these quasars are expected to be found in large-scale galaxy overdensities in the early universe \citep[e.g.,][]{Costa14}, although a large variance in the number of galaxies around SMBHs is possible as suggested by large-scale cosmological simulations \citep[e.g.,][]{Habouzit19}.

Observationally, however, testing these theories has proven to be difficult even with the largest ground-based telescopes and the {Hubble Space Telescope} (HST). Particularly, whether the most distant quasars are embedded in galaxy overdensities has been the topic of great debate.
In the past two decades, extensive efforts \citep[see][for a review]{Overzier16} have been made to search for star-forming galaxies around $z>6$ quasars \citep[e.g.,][]{Willott05, Zheng06, Kim09, Simpson14, Morselli14}. These works mostly use photometrically selected Lyman Break Galaxy (LBG) candidates as tracers. However, the large redshift uncertainty (i.e., $\Delta z\sim1$ when only a few broad-band filters are used) of photometrically-selected LBGs could significantly dilute the overdensity signal. 
Whilst spectroscopic follow-up observations of these LBGs can confirm or rule out galaxy overdensities, no consensus has been reached yet as such studies are very challenging. To date only a few $z\sim6$ quasars have been confirmed to be living in Mpc-scale (comoving) galaxy overdensities from extensive spectroscopic observations \citep{Mignoli20, Bosman20, Overzier22, Meyer22}, with the extent similar to that of protoclusters, progenitors of galaxy cluster \citep[e.g.,][]{Overzier16}.
On the other hand, ALMA has played a major role in characterizing the small-scale environment of the earliest quasars which indicates that a fraction of quasars have close companion galaxies detected in [\ion{C}{2}] emissions \citep[e.g.,][]{Decarli17,Venemans19}, however, whether overdensity of ALMA selected galaxies around quasars extends to Mpc-scale is still unclear \citep{Meyer22}.

With the launch of the {James Webb Space Telescope} (JWST), we have finally entered the era where deep spectroscopic observations of large statistical samples of high-redshift galaxies are available. Recent JWST NIRCam \citep{NIRCam} Wide Field Slitless Spectroscopic (WFSS) observations of the ultraluminous quasar J0100+2802 at $z=6.3$ \citep{Wu15} revealed that it resides in a galaxy overdensity traced by [\ion{O}{3}] emitters \citep{Kashino22}, demonstrating the transformative capability of JWST in studying the large-scale environment of high-redshift quasars. To resolve the long-standing open question of whether the earliest SMBHs reside in the most massive dark matter halos and inhabit large-scale galaxy overdensities, we designed the ASPIRE program (program ID 2078, PI: F. Wang) to search for H$\beta$+[\ion{O}{3}] emitters in the fields of 25 quasars at $z>6.5$ by utilizing the spectroscopic capabilities of NIRCam/WFSS \citep{NIRCamWFSS}. 

In this paper, we provide a brief overview of the ASPIRE project and present the discovery of a filamentary structure traced by the quasar J030516.92--315055.9 (hereafter J0305--3150) and 10 [\ion{O}{3}] emitters at $z\sim6.6$. We also report the discovery of 31 additional [\ion{O}{3}] emitters in this field with half of them clustered at $z\sim5.4$ and $z\sim6.2$.  In \S \ref{sec:obs}, we present a brief program design of ASPIRE, JWST observations and data reduction. In \S \ref{sec:o3}, we report on the line emitter search and the discovery of 41 [\ion{O}{3}] emitters in this field, and in \S \ref{sec:discuss} we discuss the properties of the filamentary structure and the complex environment of J0305--3150. Finally, we summarize our results in \S \ref{sec:summary}. Throughout the paper, we adopt a flat $\Lambda$CDM cosmology with $H_0=70~{\rm km~s^{-1}~Mpc^{-1}}$, $\rm \Omega_M=0.3$, and $\Omega_\Lambda=0.7$.

\section{Program Overview, Observations and Data Reduction} \label{sec:obs}
\subsection{ASPIRE program overview and observations}\label{sec:overview}
The science drivers of the ASPIRE program are to resolve the long-standing question of whether the earliest SMBHs reside in massive dark matter halos and inhabit galaxy overdensities, to detect the stellar light from quasar host galaxies, to understand the SMBH growth and AGN feedback, and to constrain cosmic reionization and metal enrichment in the early universe. The NIRCam/WFSS, with unprecedented spectrosocpic capabilities and remarkable imaging sensitivity and resolution, is ideal for these purposes. To put the H$\beta$+[\ion{O}{3}] lines of both quasars and physically associated galaxies at the sweet-spot of NIRCam/WFSS sensitivity, we focus on quasars at $6.5 < z < 6.8$. 
The ASPIRE program targets 25 quasars known at $6.5 < z < 6.8$ having both high-resolution (i.e., $\lesssim0\farcs5$) ALMA observations \citep[e.g.,][]{Venemans19} and high-quality optical-to-infrared spectra \citep[e.g.,][]{Yang20b,Yang21}. These quasars are selected from various on-going quasar surveys \citep[e.g.,][]{Wang19b,Yang19b,Yang20b,Banados21} and have a bolometric luminosity range of $0.5-5\times10^{47}~{\rm erg~s^{-1}}$.

Since the continua of high-$z$ galaxies are not central for our science goals and using both Grisms or observing at multiple position angles would add significant observing overheads, ASPIRE only uses Grism R for the WFSS observations. ASPIRE uses the F356W filter for the WFSS observation in the long wavelength (LW) channel and F200W filter simultaneously in the short wavelength channel (SW). This setup enables us to perform a slitless galaxy redshift survey at $\sim$3--4 $\mu$m and to obtain extremely deep imaging at 2$\mu$m. 
The resolving power of NIRCam/WFSS observation is about $R\sim1300-1600$ from $3\mu$m to $4\mu$m. The dispersion of the NIRCam/WFSS data for our configuration is about 10 \AA\ per pixel. 
The main observations are performed with a 3-point {\tt INTRAMODULEX} primary dither pattern and each primary position includes two sub-pixel dithers. This gives a survey area of $\sim$11 arcmin$^2$ for the imaging and slitless spectroscopy. 
We use the {\tt SHALLOW4} readout pattern with nine groups and one integration which gives a total on-source exposure time of 2834.5 s. To maximize the sky area coverage, both NIRCam modules are used by ASPIRE. The quasar is put at a carefully designed position ($\rm X_{offset}=-60\farcs5$, $\rm Y_{offset}=7\farcs5$) in module A to ensure that we have a full WFSS wavelength coverage and a full imaging depth for the quasar.

To identify the sources that are responsible for the slitless spectra imprinted on the NIRCam detectors, one needs to get direct imaging along with the WFSS observations. In addition, some sources that fall outside of the NIRCam imaging field of view could also produce spectra on the detectors. Therefore, the so-called out-of-field images dithering along the WFSS dispersion direction are also needed.
The direct and out-of-field imaging were performed with the same readout pattern as the main observations,  with the F115W filter in the SW and the F356W filter in the LW. The combination of direct imaging and out-of field imaging gives a total exposure time of 1417.3 s in the quasar vicinity but 472.4 s at the edge of the field for the F115W and F356W bands. We also obtained NIRISS parallel imaging which is not used here and will be presented in future work. 

In this work, we focus on one of the ASPIRE fields, centered on the quasar J0305--3150 which was observed on 2022 Aug 12 with JWST. J0305--3150 is a luminous ($L_{\rm bol}=10^{47}~{\rm erg~s^{-1}}$) quasar and hosts a SMBH with a mass of $\sim10^9~M_\odot$ at $z=6.614$ \citep{Venemans13}. High-resolution ($\sim0\farcs08\times0\farcs07$) ALMA observations of J0305--3150 indicate that it is hosted by a massive galaxy ($M_{\rm dyn}\sim3\times10^{10}~M_\odot$) with a [\ion{C}{2}]-based redshift of $z=6.6139$ \citep{Venemans19}. The ALMA observations and previous MUSE observations \citep{Farina17} identified three [\ion{C}{2}] emitters and one Lyman-$\alpha$ emitter (LAE) in the quasar vicinity, indicating that J0305--3150 could inhabit a galaxy overdensity. 
Nevertheless, \cite{Ota18} found that the narrow-band imaging selected LAEs in the J0305--3150 field show mostly an underdensity in the quasar vicinity, with a observing depth slightly shallower than that of the MUSE observation presented in \cite{Farina17}.
In addition, a HST program GO 15064 (PI: Caitlin Casey) observed J0305--3150 with ACS imaging in F606W and F814W bands and with WFC3 imaging in F105W, F125W, and F160W bands, making J0305--3150 the only ASPIRE quasar field covered by existing multi-band optical and infrared imaging from HST and an ideal target for early ASPIRE data analysis. The detailed description of the HST observations, data reduction and characterization of selected LBGs is presented in Champagne et al. submitted.

\begin{figure*}[tbh]
\centering
\includegraphics[trim=5 1 10 0, clip, width=1\linewidth]{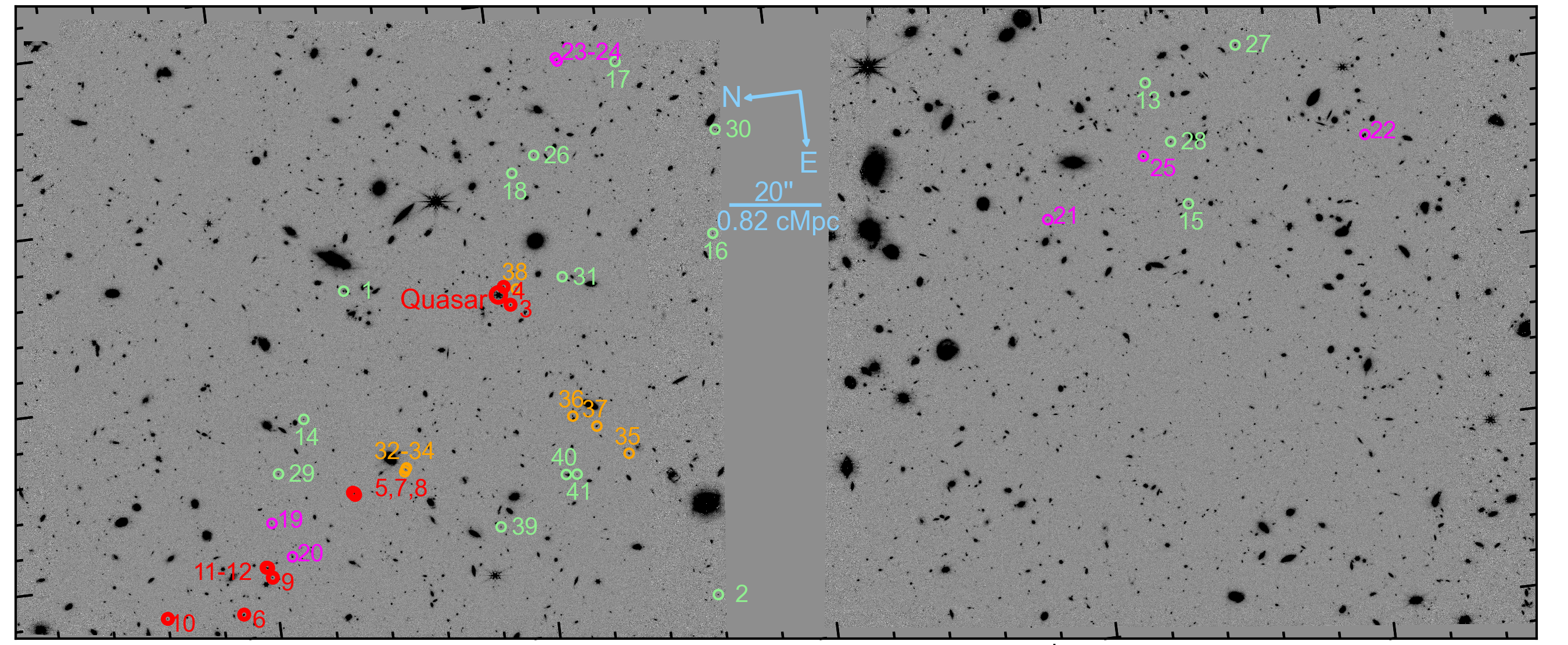}
\caption{ 
\textbf{F356W-band imaging of the J0305--3150 quasar field.}
We identified 41 [\ion{O}{3}] emitters as highlighted by colored circles. 
The quasar J0305--3150 and [\ion{O}{3}] emitters with line-of-sight velocity relative to the quasar of $\Delta v_{\rm los}<1000~{\rm km~s^{-1}}$ are highlighted by red circles. 
The magenta circles and orange circles denote the member galaxies of two galaxy overdensities at $z=6.2$ and $z=5.4$, respectively. The green circles are other field [\ion{O}{3}] emitters. The compass and the scale length (at $z=6.6$) are shown as blue lines.
}
\label{fig:map}
\end{figure*}

\subsection{NIRCam imaging data reduction}\label{sec:nircam}
We performed a careful reduction of the NIRCam images using version 1.8.3 of the JWST Calibration Pipeline\footnote{\url{https://github.com/spacetelescope/jwst}}({\tt CALWEBB}) with some additional steps as detailed below. We use the reference files ({\tt jwst\_1015.pmap}) from the version 11.16.15 of the standard Calibration Reference Data System\footnote{\url{https://jwst-crds.stsci.edu}} (CRDS)  to calibrate our data. Below, we briefly summarize our reduction procedures.

After {\tt CALWEBB} Stage 1, we masked objects and generated smooth background images for individual exposures. We  measured the {\it 1/f} noise model on a row-by-row basis and column-by-column basis for each amplifier following the algorithm proposed by \cite{Schlawin2020}. 
Then we run the Stage 2 pipeline on the {\it 1/f} noise model subtracted Stage 1 outputs. To remove extra detector-level noise features seen in the  individual Stage 2 outputs, we constructed a master median background for each combination of detector and filter based on all available exposures from the ASPIRE program. 
The master backgrounds were then scaled and subtracted for individual exposures. After that, we aligned the LW images to a reference catalog from the DESI Legacy Imaging Surveys \footnote{\url{https://www.legacysurvey.org}} \citep{Dey19} and then aligned all SW images to the calibrated LW images using {\tt tweakwcs}\footnote{\url{https://github.com/spacetelescope/tweakwcs}}. The purpose of this step is to remove astrometric offsets between different detectors, modules and filters.
The aligned and calibrated individual files were then passed through the {\tt CALWEBB} Stage 3 pipeline to create drizzled images. During the resampling step, we used a fixed pixel scale of 0\farcs0311 for SW images and 0\farcs0315 for LW images and adopted {\tt pixfrac=1}. 
The mosaiced images are further aligned to the reference catalog from the DESI Legacy Imaging Surveys for absolute astrometric calibration. Our procedure yields precise relative alignment (${\rm RMS}\simeq15$ mas) and absolute astrometric calibration (${\rm RMS}\simeq50$ mas). 
Finally, we derived and subtracted a background for each mosaic by utilizing the object detection and background estimation routines from {\tt photutils}\footnote{\url{https://photutils.readthedocs.io}} to mask out objects and estimate the background iteratively. 
We then extracted a source catalog used for spectral extraction from the fully calibrated images using {\tt SExtractor} \citep{SExtractor} with {\tt DETECT\_MINAREA=5} and {\tt DETECT\_THRESH=3}. The $5\sigma$ limiting magnitudes in 0\farcs32 diameter apertures of the calibrated images are 26.77, 27.48, and 27.85 AB magnitude for F115W, F200W, and F356W, respectively. 
To illustrate the quality of NIRCam images, we show the F356W-band mosaic in Figure \ref{fig:map}.

\begin{figure*}[tbh]
\centering
\includegraphics[trim=0 0 0 0, clip, width=0.95\linewidth]{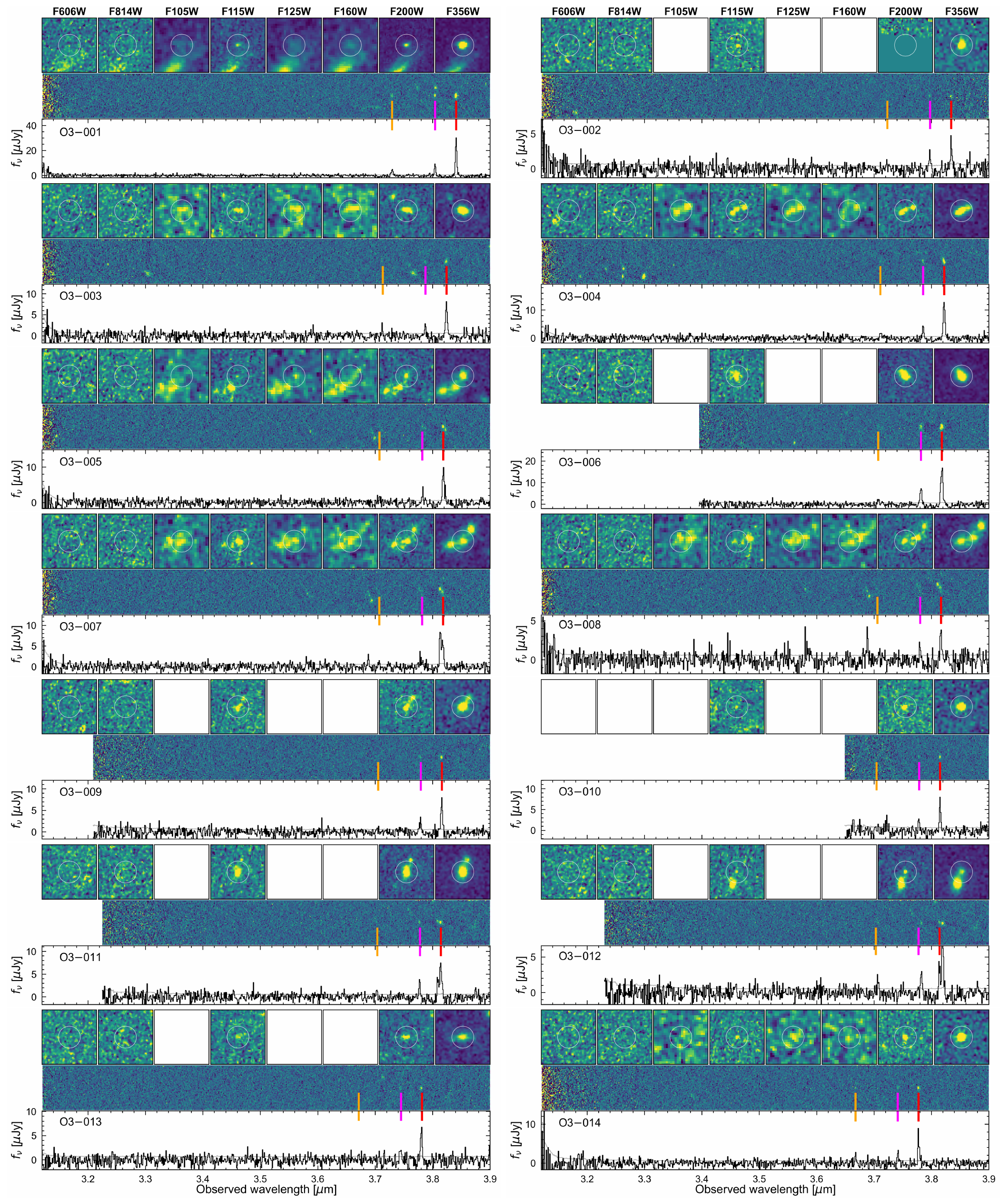}
\caption{ 
\textbf{The image cutouts and spectra of the [\ion{O}{3}] emitters.}
The cutouts from left to right show HST/F606W, HST/F814W, HST/F105W,  JWST/F115W, HST/F125W,  HST/F160W, JWST/F200W, and JWST/F356W, respectively. 
The sizes of the image cutouts are $2\farcs0\times2\farcs0$ and the radii of the white circles are 0\farcs3.
All $z>5.7$ [\ion{O}{3}] emitters are undetected in HST/F606W and HST/F814W images as expected.
The middle panels show the 2D coadded spectra and the bottom panels show the optimally extracted spectrum for each galaxy. H$\beta$, [\ion{O}{3}] $\lambda4959$ and [\ion{O}{3}] $\lambda5007$ are highlighted with orange, magenta, and red lines, respectively. The width of each 2D spectrum is 47 native pixels or $\sim3\farcs0$.
Note that we subtracted median-filtered continuum models for both 1D and 2D spectra here for visualization purposes. The spectra of the remaining [\ion{O}{3}] emitters are shown at the end of the paper. 
}
\label{fig:spec}
\end{figure*}

\subsection{NIRCam WFSS data reduction}\label{sec:wfss}
We used the {\tt CALWEBB} Stage 1 pipeline to calibrate the detector-level signals and ramp fitting for individual NIRCam WFSS exposures. After this step, we subtracted the {\it 1/f} noise pattern using the same routines that are used for the NIRCAM direct image processing, but we only subtracted the stripes along columns since the spectra are dispersed along rows for Grism-R. We then performed flat-fielding of the WFSS exposures using direct imaging flat reference files since the WFSS flat reference files are not available. We also assigned {\tt WCS} information for individual files based on the distortion reference files in CRDS using the {\tt assign\_wcs} routine in the {\tt CALWEBB} pipeline. 
We measured astrometric offsets between each of the SW images (i.e., the exposures taken simultaneously with WFSS exposures) and the fully calibrated F356W-band mosaic. The measured offsets were applied to the spectral tracing model in a later stage when tracing objects. All further data processing steps after this stage were performed using our custom scripts as detailed below.

In the standard {\tt CALWEBB} pipeline, the WFSS background subtraction is performed by scaling the theoretical background to that observed in each individual integration. However, the accuracy of the theoretical background has not been fully characterized because of limited existing in-flight calibrations. To resolve this issue, we constructed median background models based on all ASPIRE observations obtained at similar times, which were scaled and subtracted from individual WFSS exposures. 

Before extracting spectra from WFSS observations, we need to construct the spectral tracing models, dispersion models and sensitivity functions. The methodology used for constructing these models is described in detail in \cite{Sun22b}. In this work, we used a spectral tracing model constructed using the spectral traces of point sources observed in the LMC field (PID \#1076) which has better detector coverage than that used in \cite{Sun22b} and \cite{Sun22a}, as well as updated sensitivity functions derived from Cycle-1 calibration programs (PIDs \#1536, \#1537, \#1538). The sensitivity functions measured from ERO calibration and Cycle-1 calibration are consistent with each other to an accuracy of better than 2\% over most wavelength coverage. We note that the dispersion and tracing models are measured from a different filter configuration and could have a small shift after considering different filter offsets. In the spatial direction (along columns for Grism-R), we measured a half-pixel offset between the model and our data. We cannot measure the offset along the dispersion direction (along rows for Grism-R) which requires in-flight observation of a wavelength calibrator and is not available yet. We, however, expect that the constant offset should be small ($<100~{\rm km~s^{-1}}$ or $\Delta z < 0.003$ for [\ion{O}{3}] emitters) since the offset in the spatial direction is smaller than one pixel. Additionally, the redshift of one of the [\ion{O}{3}] emitters is consistent with the galaxy's Ly$\alpha$ redshift ($z_{\rm [OIII]}-z_{\rm Ly \alpha}=0.002$; see \S \ref{sec:vicinity}) and the redshift of the quasar derived from [\ion{O}{3}] is also consistent with the quasar's [\ion{C}{2}] redshift ($z_{\rm [OIII]}-z_{\rm [CII]}=0.001$; Yang et al. submitted). 
Such a small wavelength zeropoint offset does not affect any scientific results presented here.

With the models described above, we then extracted both 2D spectra and 1D spectra for all sources ($\sim5000$) detected in the F356W direct imaging (\S \ref{sec:nircam}). We extracted two versions of spectra for each object. For the first version, we extracted the 1D spectra from individual WFSS exposures using both boxcar and optimal extraction algorithms \citep{Horne86}. The extracted individual 1D spectra were then combined with inverse variance weighting and outlier rejection. For the second version, we first extracted 2D spectra for each source from each individual exposure and then stacked the 2D spectra after registering them to a common wavelength and spatial grid following the histogram2D technique implemented in the PypeIt software \citep{Pypeit} to avoid interpolations. We then extracted 1D spectra from the stacked 2D spectra using both boxcar and optimal extraction algorithms. 
We used an aperture diameter of five pixels (0\farcs315) for the boxcar spectral extraction. The pixel scales for both 1D and 2D spectra are resampled to be 10 \AA\ per pixel.
During the 1D spectral extraction process for both methods, we performed iterative background subtraction to subtract off background residuals. 
We note that the two spectra sets agree well with each other, and we only used the second approach in this work by considering that the profile fitting used for optimal extraction is slightly better characterized in the stacked 2D spectra for very faint galaxies. 
In the highest sensitivity regions of the slitless spectra, 
the $5\sigma$ emission line detection limit for a point source is estimated to be $2.0\times10^{-18}~{\rm erg~s^{-1}~cm^{-2}}$ by integrating the spectra over 50 \AA\ (i.e., $\sim2\times$ spectral resolution). This corresponds to a $5\sigma$ line luminosity limit of $9.9\times10^{41}~{\rm erg~s^{-1}}$ at $z=6.6$.

\begin{deluxetable*}{cccccc}
\tablecaption{The list of identified [\ion{O}{3}] emitters.}
\tabletypesize{\scriptsize}
\tablewidth{0pt}
\tablehead{
\colhead{Name\tablenotemark{a}} &
\colhead{RA} &
\colhead{Decl.} &
\colhead{$z$\tablenotemark{b}} &
\colhead{$L_{\rm [O\,{\sc III}]4960,5008}$} &
\colhead{$\rm EW_{[O\,{\sc III}]4960,5008}$} 
}
\startdata
                        &              &        &      & $\rm 10^{42}~erg ~s^{-1}$ &  \AA\     \\
\hline
ASPIRE-J0305M31-O3-001 & 03:05:16.530 & -31:50:22.663 & 6.669 & $12.39\pm0.63$ & $679\pm35$ \\
ASPIRE-J0305M31-O3-002 & 03:05:22.444 & -31:51:35.268 & 6.656 & $2.05\pm0.41$ & $500\pm100$ \\
ASPIRE-J0305M31-O3-003 & 03:05:17.109 & -31:50:58.253 & 6.636 & $3.87\pm0.39$ & $550\pm56$ \\
ASPIRE-J0305M31-O3-004 & 03:05:16.793 & -31:50:57.264 & 6.631 & $6.52\pm0.47$ & $468\pm34$ \\
ASPIRE-J0305M31-O3-005 & 03:05:19.954 & -31:50:19.274 & 6.624 & $4.27\pm0.59$ & $201\pm28$ \\
ASPIRE-J0305M31-O3-006 & 03:05:21.785 & -31:49:52.576 & 6.623 & $10.72\pm0.89$ & $600\pm50$ \\
ASPIRE-J0305M31-O3-007 & 03:05:19.971 & -31:50:19.589 & 6.623 & $2.94\pm0.56$ & $91\pm17$ \\
ASPIRE-J0305M31-O3-008 & 03:05:19.996 & -31:50:19.682 & 6.621 & $1.94\pm0.44$ & $78\pm18$ \\
ASPIRE-J0305M31-O3-009 & 03:05:21.221 & -31:49:59.745 & 6.619 & $3.56\pm0.56$ & $528\pm83$ \\
ASPIRE-J0305M31-O3-010 & 03:05:21.697 & -31:49:36.022 & 6.617 & $2.91\pm0.29$ & $1523\pm153$ \\
ASPIRE-J0305M31-O3-011 & 03:05:21.048 & -31:49:58.981 & 6.616 & $4.50\pm0.73$ & $189\pm31$ \\
ASPIRE-J0305M31-O3-012 & 03:05:21.040 & -31:49:58.662 & 6.615 & $1.45\pm0.48$ & $110\pm36$ \\
ASPIRE-J0305M31-O3-013 & 03:05:14.691 & -31:53:20.922 & 6.550 & $2.58\pm0.68$ & $404\pm107$ \\
ASPIRE-J0305M31-O3-014 & 03:05:18.612 & -31:50:10.651 & 6.542 & $2.59\pm0.63$ & $718\pm175$ \\
ASPIRE-J0305M31-O3-015 & 03:05:16.825 & -31:53:27.012 & 6.506 & $6.07\pm1.20$ & $1331\pm264$ \\
ASPIRE-J0305M31-O3-016 & 03:05:16.326 & -31:51:43.739 & 6.497 & $3.09\pm0.63$ & $678\pm140$ \\
ASPIRE-J0305M31-O3-017 & 03:05:13.220 & -31:51:27.269 & 6.496 & $2.08\pm0.47$ & $580\pm133$ \\
ASPIRE-J0305M31-O3-018 & 03:05:14.891 & -31:51:02.045 & 6.419 & $2.05\pm0.25$ & $751\pm92$ \\
ASPIRE-J0305M31-O3-019 & 03:05:20.302 & -31:50:01.008 & 6.298 & $2.30\pm0.45$ & $718\pm141$ \\
ASPIRE-J0305M31-O3-020 & 03:05:20.911 & -31:50:04.545 & 6.298 & $7.22\pm0.58$ & $392\pm32$ \\
ASPIRE-J0305M31-O3-021 & 03:05:16.800 & -31:52:56.271 & 6.292 & $1.77\pm0.51$ & $860\pm250$ \\
ASPIRE-J0305M31-O3-022 & 03:05:16.025 & -31:54:06.873 & 6.280 & $8.41\pm1.30$ & $253\pm39$ \\
ASPIRE-J0305M31-O3-023 & 03:05:13.031 & -31:51:14.492 & 6.262 & $1.23\pm0.38$ & $456\pm141$ \\
ASPIRE-J0305M31-O3-024 & 03:05:13.088 & -31:51:14.824 & 6.257 & $6.05\pm0.61$ & $835\pm84$ \\
ASPIRE-J0305M31-O3-025 & 03:05:15.928 & -31:53:18.545 & 6.253 & $6.59\pm1.00$ & $741\pm113$ \\
ASPIRE-J0305M31-O3-026 & 03:05:14.633 & -31:51:07.238 & 6.089 & $1.92\pm0.34$ & $864\pm154$ \\
ASPIRE-J0305M31-O3-027 & 03:05:14.240 & -31:53:41.306 & 6.081 & $3.10\pm1.08$ & $161\pm56$ \\
ASPIRE-J0305M31-O3-028 & 03:05:15.738 & -31:53:24.841 & 6.078 & $2.79\pm0.59$ & $269\pm57$ \\
ASPIRE-J0305M31-O3-029 & 03:05:19.481 & -31:50:03.737 & 5.957 & $1.84\pm0.42$ & $373\pm85$ \\
ASPIRE-J0305M31-O3-030 & 03:05:14.575 & -31:51:46.999 & 5.815 & $4.20\pm0.44$ & $2316\pm247$ \\
ASPIRE-J0305M31-O3-031 & 03:05:16.745 & -31:51:10.141 & 5.662 & $1.92\pm0.46$ & $2758\pm670$ \\
ASPIRE-J0305M31-O3-032 & 03:05:19.648 & -31:50:31.465 & 5.448 & $3.06\pm0.45$ & $448\pm66$ \\
ASPIRE-J0305M31-O3-033 & 03:05:19.666 & -31:50:31.308 & 5.447 & $1.95\pm0.52$ & $253\pm68$ \\
ASPIRE-J0305M31-O3-034 & 03:05:19.709 & -31:50:30.986 & 5.442 & $10.43\pm0.92$ & $1922\pm172$ \\
ASPIRE-J0305M31-O3-035 & 03:05:19.870 & -31:51:19.826 & 5.432 & $4.14\pm0.49$ & $378\pm44$ \\
ASPIRE-J0305M31-O3-036 & 03:05:19.122 & -31:51:08.694 & 5.432 & $2.38\pm0.32$ & $203\pm28$ \\
ASPIRE-J0305M31-O3-037 & 03:05:19.340 & -31:51:13.606 & 5.432 & $4.32\pm0.64$ & $339\pm51$ \\
ASPIRE-J0305M31-O3-038 & 03:05:16.856 & -31:50:59.955 & 5.428 & $3.30\pm0.51$ & $491\pm77$ \\
ASPIRE-J0305M31-O3-039 & 03:05:20.843 & -31:50:50.277 & 5.399 & $2.74\pm0.43$ & $1228\pm194$ \\
ASPIRE-J0305M31-O3-040 & 03:05:20.095 & -31:51:05.726 & 5.395 & $3.09\pm1.11$ & $156\pm56$ \\
ASPIRE-J0305M31-O3-041 & 03:05:20.113 & -31:51:08.072 & 5.341 & $3.24\pm1.34$ & $4115\pm1706$
\enddata
\tablenotetext{a}{Galaxies spectroscopically confirmed from the ASPIRE program will be named as ASPIRE-JHHMMXDD-TT-NNN, where JHHMMXDD represents quasar field name, TT denotes galaxy type (i.e., O3 for [\ion{O}{3}] emitters, HA for H$\alpha$ emitters and so on) and NNN represents galaxy ID number.}
\tablenotetext{b}{The redshift uncertainties of galaxies are dominated by systematic wavelength zeropoint offsets of current spectral dispersion model which would be improved with more JWST in-flight calibrations. To be conservative, we budget a $\Delta z=0.003$ for all galaxies identified in this work.}
\label{tbl:emitters}
\end{deluxetable*}

\begin{figure}[tbh]
\centering
\includegraphics[trim=0 0 0 0, clip, width=1.0\linewidth]{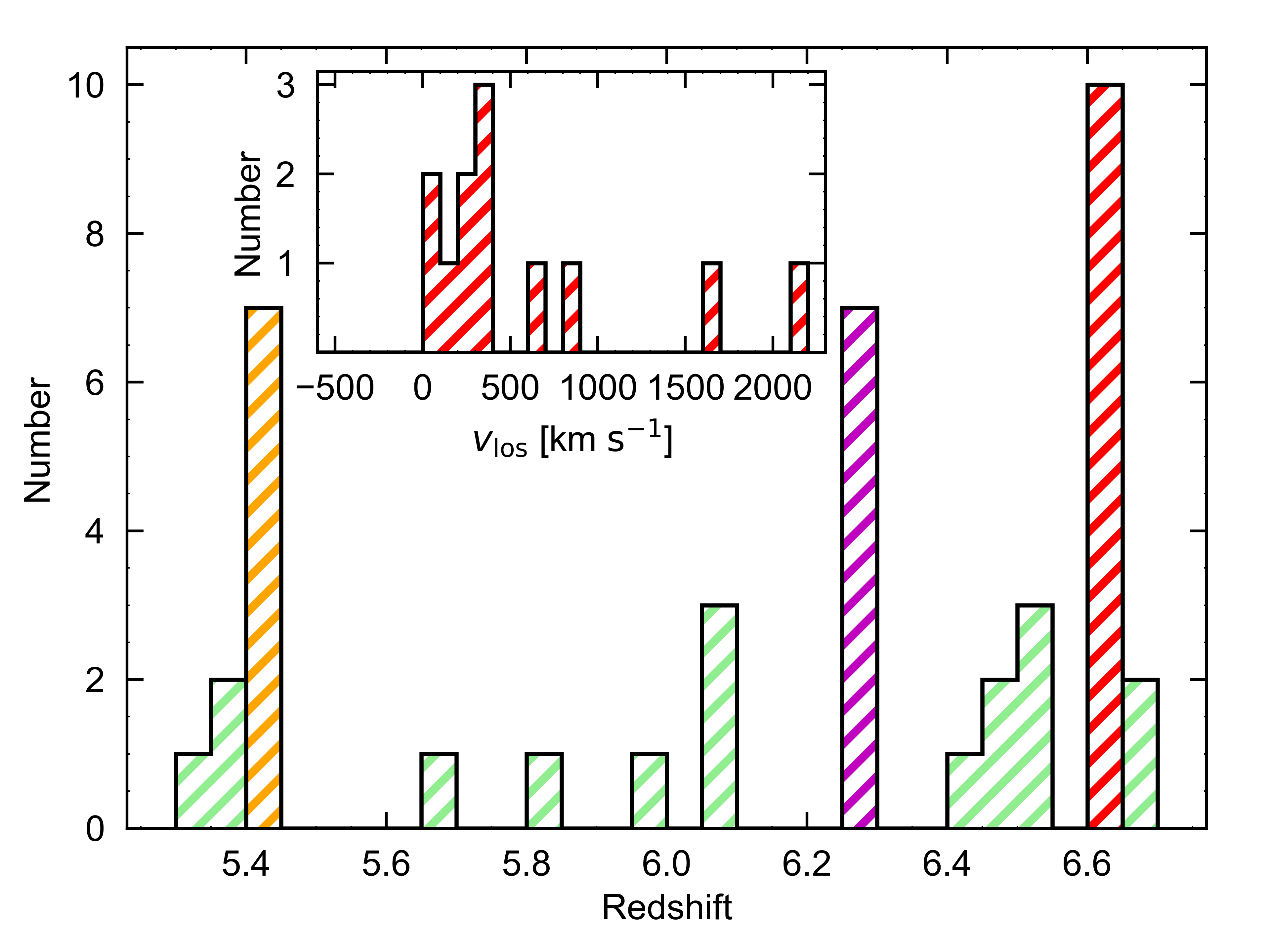}
\caption{ 
\textbf{Redshift distribution of [\ion{O}{3}] emitters.}
The [\ion{O}{3}] emitters are strongly clustered in redshift space. There are obvious galaxy number excess at $z\sim5.4$, $z\sim6.2$, and $z\sim6.6$ in this field. 
The most significant excess in galaxy number appears at $z\sim6.6$, the same redshift as the quasar. The inner plot shows the distribution of the line-of-sight velocities relative to the quasar of galaxies at $6.6<z<6.7$. Ten of the twelve galaxies at $z\sim6.6-6.7$ have $v_{\rm los}<1000~{\rm km~s^{-1}}$ relative to the quasar. A positive velocity offset means that the galaxy has a higher redshift than the quasar. Limited by the precision of NIRCam/WFSS wavelength calibration, the redshifts of [\ion{O}{3}] emitters could have a constant offset up to 0.003 as discussed in \S \ref{sec:wfss}, which could explain the asymmetric distribution of $v_{\rm los}$.
}
\label{fig:redshift}
\end{figure}

\section{Discovery of 41 [\ion{O}{3}] emitters at $5.3<\lowercase{z}<6.7$} \label{sec:o3}
Directly detecting the rest-frame optical [\ion{O}{3}] $\lambda\lambda$4960\AA, 5008\AA\ emission lines from EoR galaxies had been impossible until the launch of JWST. Within the first JWST observations, there are five [\ion{O}{3}] emitters identified at $z>5$ from NIRSpec observations of the SMACS0723 field \citep[e.g.,][]{Carnall23, Curti23} and another four galaxies discovered from NIRCam WFSS observations of the P330-E standard star field \citep{Sun22b, Sun22a}. The more recent deep NIRCam WFSS observation of the J0100+2802 quasar field identified more than 100 [\ion{O}{3}] emitters at $z>5$, demonstrating the power of the NIRCam WFSS mode \citep{Kashino22}. 
These [\ion{O}{3}] emitters are powered by young stellar populations and have high H$\beta$+[\ion{O}{3}] equivalent widths (EW), moderately low metallicities, and high ionization states 
\citep[e.g.,][]{Matthee22, Sun22b}.

To search for [\ion{O}{3}] emitters in the J0305--3150 field, we developed a set of scripts to automatically search for line emitters from the extracted 1D spectra. We first produced a median-filtered continuum model with a window size of 51 pixels (510 \AA), which is subtracted from the optimally extracted spectrum. We then applied a peak finding algorithm to search for all peaks with S/N$>$3 and rejected peaks with the two nearest pixels having S/N$<$1.5 which are mostly residuals from hot pixels or cosmic rays. For the remaining detected peaks, we performed a Gaussian fitting to measure the full-width-at-half-maximum (FWHM) and S/N, where the S/N was measured directly from the continuum-subtracted spectrum by integrating pixels within $\pm1\times$ FWHM. In this step, we rejected lines with S/N$<5$ and FWHM widder than seven times the spectral resolution ($R\sim25$ \AA, or about 2.5 pixels) or FWHM narrower than half of the spectral resolution to further reject fake lines.
For each detected emission line, we first assumed that it is the [\ion{O}{3}]$\lambda$5008\AA\ line and then measured the S/N of the [\ion{O}{3}]$\lambda$4960\AA\ line at the expected wavelength by assuming that it has the same FWHM as [\ion{O}{3}]$\lambda$5008\AA. We identified an object as an [\ion{O}{3}] emitter candidate only if there is a $>2\sigma$ significance line detection at the expected wavelength. Finally, we visually inspected the spectra for all [\ion{O}{3}] emitter candidates and identified 41 [\ion{O}{3}] emitters in the J0305--3150 field in addition to the quasar. 
To check the robustness of our line emitter searching algorithm, we further visually inspected all extracted spectra ($\sim 5000$ sources in total). From the visual inspection, we found that our automatic line emitter searching method recovers all the [\ion{O}{3}] emitters identified by eye.

The basic information of these [\ion{O}{3}] emitters are listed in Table \ref{tbl:emitters}. 
These [\ion{O}{3}] emitters span a redshift range of $5.3<z<6.7$, a [\ion{O}{3}] luminosity range of $L_{[\rm O\,{\sc III}]4960,5008} {\rm = (1.2-12.4)\times10^{42}~erg~s^{-1}}$, and a rest-frame equivalent width range of $\rm 80~\AA \lesssim EW_{[O\,{\sc III}]4960,5008} \lesssim 4100$ \AA.
The $L_{[\rm O\,{\sc III}]4960,5008}$ was determined from Gaussian fitting of both [\ion{O}{3}]$\lambda5008$\AA\ and [\ion{O}{3}]$\lambda4960$\AA\ lines and the $\rm EW_{[O\,{\sc III}]4960,5008}$ was estimated based on the H$\beta$+[\ion{O}{3}] line fitting result and F356W-band photometry by assuming that $f_\lambda$ in the continuum is a constant across the F356W passband.
The positions of these galaxies are highlighted in Figure \ref{fig:map} and the JWST and HST imaging thumbnails and the extracted spectra of these galaxies are presented in Figure \ref{fig:spec}. In Figure \ref{fig:redshift}, we show the redshift distribution of all [\ion{O}{3}] emitters identified in the field. 
Obviously, most of these [\ion{O}{3}] emitters are clustered at $z\sim5.4$ (orange), $z\sim6.2$ (magenta), and $z\sim6.6$ (red). This indicates that we are seeing three galaxy overdensities in just one ASPIRE field. In particular, the galaxy overdensity at $z\sim6.6$ coincides with the redshift of the quasar and has the most abundant galaxies, indicating that quasar J0305--3150 traces a significant galaxy overdensity. In the following sections, we will focus on the galaxy overdensity at $z\sim6.6$, more detailed SED modeling and characterization of these [\ion{O}{3}] emitters will be presented in a subsequent paper (Champagne et al. in prep). 

\section{Discussion} \label{sec:discuss}
\begin{figure*}[tbh]
\centering
\includegraphics[trim=30 30 0 50, clip, width=0.57\linewidth]{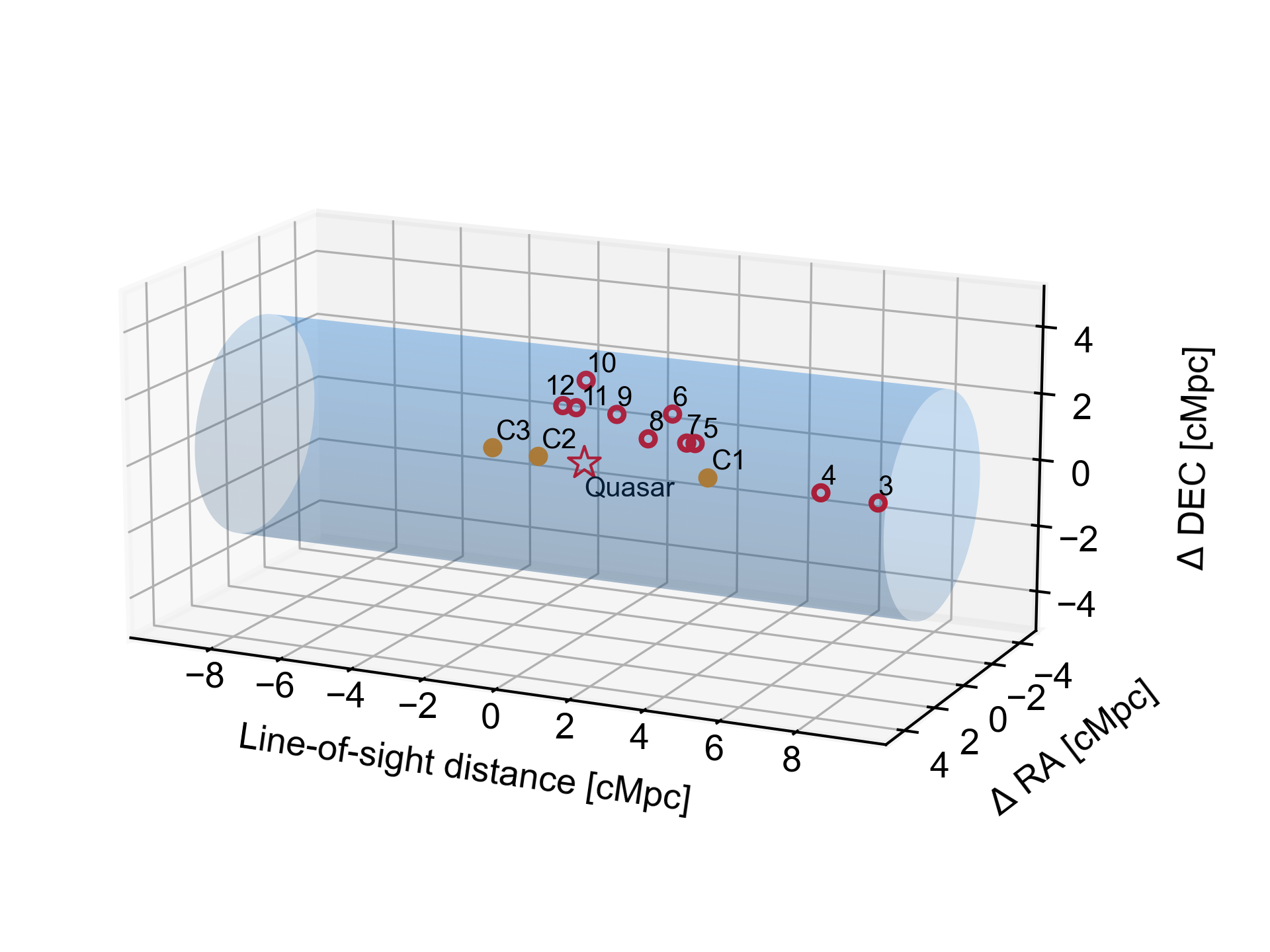}
\includegraphics[trim=0 0 50 0, clip, width=0.42\linewidth]{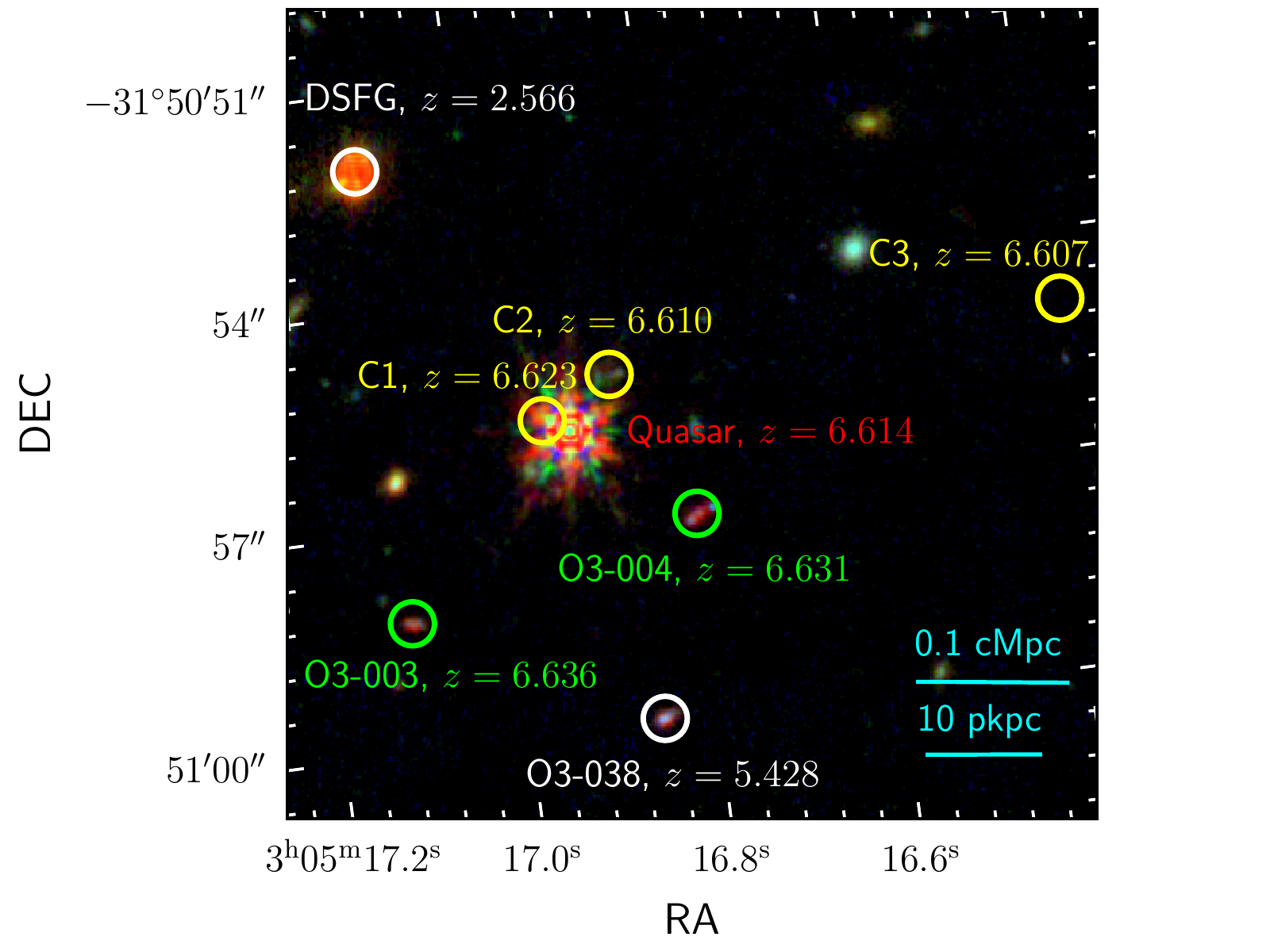}
\caption{ 
\textbf{Left: 3D structure of the galaxy overdensity at $z=6.6$.}
In this plot, we show both [\ion{O}{3}] emitters and [\ion{C}{2}] emitters in the vicinity of the quasar. The blue shaded region highlights a cylinder volume with line-of-sight length of 2000 km $\rm s^{-1}$ (or 18.8 cMpc) and a projected radius of 3.28 cMpc at $z=6.6$ (corresponding to an effective area of $\rm 5.5~ arcmin^2$ in the projected plane). 
\textbf{Right: The immediate vicinity of quasar J0305--3150.}
The background is a RGB image made using NIRCam imaging in F115W (B), F200W (G), and F356W (R). 
Quasar J0305--3150 is surrounded by three [\ion{C}{2}] emitters \citep[C1, C2, and C3;][]{Venemans19} and two [\ion{O}{3}] emitters (ASPIRE-J0305M31-O3-003 and ASPIRE-J0305M31-O3-004) with line-of-sight velocities relative to the quasar of $\Delta v_{\rm los}<1000~{\rm km~s^{-1}}$. Galaxy C3 is undetected in our deep JWST observations which indicates that it is a dusty star forming galaxy (DSFG). Galaxy [\ion{O}{3}]-04 is also detected in Ly$\alpha$ \citep{Farina17}. Two foreground galaxies are also shown with a DSFG at $z=2.566$ and a [\ion{O}{3}] emitter at $z=5.428$.
}
\label{fig:vicinity}
\end{figure*}

\subsection{An overdense filamentary structure around quasar J0305--3150}\label{sec:structure}
Our new JWST observations allow us to characterize the Mpc-scale (comoving) environment of the $z=6.6$ quasar J0305--3150 for the first time. Among the 41 [\ion{O}{3}] emitters identified in the J0305--3150 field, there are twelve [\ion{O}{3}] emitters (emitters 1--12) at $z\sim6.61-6.67$. In particular, ten of them (emitters 3-12) have line-of-sight velocity separations from the quasar with $\Delta v_{\rm los}<1000~{\rm km~s^{-1}}$ (or $<9.4$ cMpc; see the insert panel of Figure \ref{fig:redshift}), with a projected distance of $10-550$ proper kpc or $80-4100$ comoving kpc relative to the quasar. The velocity dispersion along the line-of-sight of these ten [\ion{O}{3}] emitters is $\sigma_v=250~{\rm km~s^{-1}}$, similar to that found in the overdensity traced by the quasar J0100+2802 at a lower redshift \citep{Kashino22}. Additionally, these ten [\ion{O}{3}] emitters are distributed along a filamentary structure towards the luminous quasar.
In the left panel of Figure \ref{fig:vicinity}, we show the zoom-in 3D structure of this galaxy overdensity. The filamentary structure extends at least 4.1 cMpc along the projected plane and 8.0 cMpc along the line-of-sight direction. This is consistent with the size of protoclusters seen in cosmological simulations \citep[e.g.,][]{Overzier09}. 

To characterize the galaxy overdensity of this structure, we first measure its galaxy number density using a cylinder volume. The projected sky area is estimated to be $5.5~{\rm arcmin}^2$ (or $33.8~{\rm cMpc}^2$) at $z\sim6.6$ in module A and the length of the cylinder is estimated by assuming the line-of-sight range of $\pm1000~{\rm km~s^{-1}}$ which delivers a cylinder volume of 637 cMpc$^3$. Such cylinder is highlighted by the shaded region in the left panel of Figure \ref{fig:vicinity}. Using the ten [\ion{O}{3}] emitters, we measure the number density of the [\ion{O}{3}] emitters in this structure to be $n_{\rm gal}= 10^{-1.80^{+0.15}_{-0.16}}~{\rm cMpc^{-3}}$.
Since module A has a better sensitivity than that of module B and all [\ion{O}{3}] emitters in the structure traced by the quasar are within module A (see Figure \ref{fig:map}), we measure the average galaxy number density using the 24 field galaxies in module A (i.e., after excluding ten [\ion{O}{3}] emitters in the structure traced by the quasar). The effective survey area of NIRCam/WFSS observations depends on the wavelength of the line and thus the redshift of a given line emitter. We follow the method used by \cite{Sun22a} to estimate the survey volume from $z=5.2$ to $z=7.0$ at a step of $\Delta z = 0.1$ and measure that the effective survey volume of module A to be 20813 cMpc$^3$ after excluding the volume of the cylinder described above. We then measure a volume density of field galaxies to be $ \bar{n}_{\rm gal}= 10^{-2.94^{+0.10}_{-0.10}}~{\rm Mpc^{-3}}$. These estimates indicate that the structure traced by the quasar has a galaxy overdensity of $\delta_{\rm gal} = \frac{n_{\rm gal}}{\bar{n}_{\rm gal}}-1=12.6^{+5.9}_{-5.0}$ within a cylinder volume of 637 cMpc$^3$. Note that we used a cylinder with the length of $\pm1000~{\rm km~s^{-1}}$ (or 18.8 cMpc) when estimating the galaxy overdensity. If we use the length (i.e., 8 cMpc) of the structure as determined by the ten [\ion{O}{3}] emitters, we would expect to have a galaxy overdensity close to 30 within a smaller volume. 
Additionally, the field galaxy density could be overestimated given that this field contains two galaxy overdensities at $z=5.4$ and $z=6.2$, respectively.
These estimates indicate that J0305--3150 traces one of the most overdense structure known in the early Universe \citep[e.g.,][]{Overzier22}.

To better understand this galaxy overdensity, we compare galaxy number counts with cosmological simulations. Using `zoom-in' cosmological simulations, \citet{Costa14} and \citet{vanderVlugt19} show that SMBHs with $M_{\rm BH} \sim 10^9 \, \rm M_\odot$ form by $z \, = \, 6$ as long as they are hosted by massive haloes with virial masses $\approx (3 \-- 5) \times 10^{12} \, \rm M_\odot$. We focus here on a suite of identical simulations targeting six such haloes, performed at $8$ times better mass resolution than \citet{Costa14}, such that the smallest resolved haloes have a virial mass $\approx 3 \times 10^7\, \rm M_\odot$. The masses of the black holes hosted by the most massive galaxies in these haloes range from $10^9 \, \rm M_\odot$ to  $10^{10} \, \rm M_\odot$ at $z \, = \, 6$.
Following the empirical [\ion{O}{3}]$-M_{\rm UV}$ relation presented in \citet{Matthee22}, we compute [\ion{O}{3}] luminosities for all galaxies in the simulation at $z \, = \, 6.6$ and then evaluate the total number of satellite galaxies with an unobscured [\ion{O}{3}] luminosity $L_{\rm [OIII]\lambda5008\AA} > 9.9 \times 10^{41} \rm erg \, s^{-1}$ 
($5\sigma$ limit of our observation). Within a circular region of $33.8 \, \rm cMpc^2$, the satellite number count varies significantly, ranging from $\approx 10$ to $\approx 36$, with a mean of $21.5$ across the halo sample. While our lower number count estimates are consistent with the number of satellites observed in this study, the simulations typically predict a factor $2$ higher number of satellites than found for J0305--3150. While this small discrepancy could indicate that the models overestimate the quasar halo mass, it could also be resolved if several of these satellites are significantly dust-obscured or winds driven by stellar feedback are stronger than envisaged in the simulations \citep[see][]{Costa14}. 
Note that the virial masses of the six haloes discussed here all lie in the range $10^{14} \-- 10^{15} \, \rm M_\odot$  at $z \, = \, 0$, such that they all represent protocluster environments at $z \, = \, 6$. This suggests that the structure traced by J0305--3150 is consistent with an protocluster environment in the early Universe and could eventually collapse into a massive galaxy cluster.

Furthermore, \cite{Habouzit19} investigated the galaxy number counts around high-redshift quasars with Horizon-AGN. Limited by the volume, the biggest SMBH they can find in the simulation at $z\sim6$ is a $10^{8.3}~M_\odot$ SMBH that is hosted by a massive galaxy with a stellar mass of $M_{\star}=10^{10.4}~M_\odot$.
In the field of this SMBH, they identified six (five) $M_{\star}\geqslant 10^8~M_\odot$ galaxies with a projected length of 50 cMpc (10 cMpc) and within the survey area of our interests (5.5 arcmin$^2$ or 33.8 cMpc$^2$). Note that, the 5$\sigma$ limit of our [\ion{O}{3}] line flux corresponding to $M_{\star}\simeq10^{8}~M_\odot$ according to the [\ion{O}{3}]$-M_{\rm UV}$ relation presented in \citet{Matthee22} and the $M_{\rm UV}-M_{\star}$ relation at $z\sim7$ from \cite{Stefanon21}, though both relations have large scatters.
We confirmed these numbers by counting galaxies ($M_{\star}\geqslant 10^{8}\, \rm M_{\odot}$) in a cylinder of length 18.8 cMpc and projected area of 33.8 cMpc$^2$ (see Fig.~\ref{fig:vicinity}) around the most massive SMBHs produced in the other cosmological simulations. At $z=6$, the few SMBHs with $M_{\star}\geqslant 10^8~M_\odot$ are surrounded by a median of $3.5-4$ galaxies (mean: 3.5-10.3) in SIMBA \citep{Dave19} and EAGLE \citep{Schaye15}. These numbers are lower than the number of galaxies identified in J0305--3150 field, but this is not surprising as the quasar is powered by a more massive SMBH of $10^{9}\, \rm M_{\odot}$. BlueTides \citep{Feng16,DiMatteo17} is the only current large-scale cosmological simulation with enough volume to produce several (7) SMBHs with $\sim 10^{9}\, \rm M_{\odot}$. At $z=7$ (the last output), these SMBHs are surrounded by a median of 15 galaxies with $M_{\star}\geqslant 10^{8}\, \rm M_{\odot}$ (from 11 to 66 galaxies).
This is in good agreement with our present observation of J0305--3150's environment. We caution, however, that all the simulations of the field produce a variety of environments around SMBHs at high redshift. To increase statistics, we repeated the exercise with less massive SMBHs of $\sim 10^{7.5}\, \rm M_{\odot}$ and found a large diversity in the number of surrounding galaxies with a median ranging from 2 to 21 (mean from 3.2 to 20.1) when considering Horizon-AGN \citep{Dubois14}, Illustris \citep{Vogelsberger14}, TNG100 \citep{Pillepich18}, TNG300, EAGLE, SIMBA, Astrid \citep{Bird22,Ni22}, and BlueTides. These numbers significantly decrease when only galaxies with $M_{\star}\geqslant 10^{8.5}$ or $10^{9}\, \rm M_{\odot}$ are detectable in the SMBH field of view, making the detection of overdensities particularly challenging.

These comparisons indicate that the most massive BHs in cosmological simulations are usually traced by galaxy overdensities but with a broad range of galaxy numbers. In addition, shallow observations (i.e., only sensitive to galaxies with $M_\star\gtrsim10^9\, \rm M_{\odot}$) could easily miss galaxy overdensities limited by the number of galaxies in the quasar fields. The completion of the ASPIRE program will enable a more systematic comparison between a large sample of quasar fields and the cosmological simulations by investigating galaxy numbers, galaxy velocity distribution and quasar properties, which will deliver a more comprehensive understanding of the environment of the earliest SMBHs.

\subsection{The complex vicinity of quasar J0305--3150} \label{sec:vicinity}
Deep ALMA observations of J0305--3150 uncovered three [\ion{C}{2}]158$\mu$m emitters in the quasar vicinity: galaxy C1 with a [\ion{C}{2}] luminosity of $L_{\rm [CII]}=(4.7\pm0.5)\times10^8~L_\odot$ at a projected distance of 2.0 pkpc, C2 with $L_{\rm [CII]}=(3.4\pm0.8)\times10^8~L_\odot$ at a projected distance of 5.1 pkpc, and C3 with $L_{\rm [CII]}=(1.2\pm0.2)\times10^9~L_\odot$ and a far-infrared luminosity of $L_{\rm FIR}=(1.3\pm0.3)\times10^{12}~L_\odot$ at a projected distance of 37.4 pkpc \citep{Venemans19}. The positions of these galaxies are highlighted in the right panel of Figure \ref{fig:vicinity}, where we show the deep NIRCam imaging of the quasar vicinity (i.e., $\sim60\times60$ pkpc$^2$ in the projected plane). 
Galaxy C1 is too close ($r<0\farcs4$) to the quasar and cannot be detected by JWST without careful PSF modeling, which will be discussed in a future work (Yang et al. in prep). Galaxy C2 is detected by NIRCam in all three bands though the flux is contaminated by the PSF spikes of the bright quasar. This galaxy is too faint to be detected in the WFSS observation. Galaxy C3 is remarkably bright in both [\ion{C}{2}] and dust continuum ($f_{\rm 1mm}=0.58\pm0.12$ mJy) but undetected in our deep NIRCam images with F115W ($f_{1.15\mu {\rm m},3\sigma}<8.9$ nJy), F200W ($f_{2.0\mu {\rm m},3\sigma}<4.1$ nJy), and F356W ($f_{3.56\mu {\rm m},3\sigma}<2.6$ nJy), which indicates that C3 is a dust obscured galaxy similar to optically invisible dusty star-forming galaxies (DSFG) found at lower redshifts \citep{WangT19}.
We also note that the quasar's position measured from the NIRCam imaging is consistent with that measured from the ALMA observation with an offset of $0\farcs02$ (i.e., less than one third of the resolution of both NIRCam and ALMA).

In addition to the three [\ion{C}{2}] emitters, our JWST observations detected two $z\sim6.6$ [\ion{O}{3}] emitters in the quasar immediate vicinity. Galaxy O3-003 has a projected distance of 18 pkpc and $v_{\rm los}=870~{\rm km~s^{-1}}$ relative to the quasar and O3-004 has a projected distance of 11 pkpc and $v_{\rm los}=670~{\rm km~s^{-1}}$. These two galaxies have [\ion{O}{3}]$\lambda$5008\AA\ luminosities of $L_{\rm [OIII]}=(3.0\pm0.2)\times10^{42}~{\rm erg~s^{-1}}$ and $L_{\rm [OIII]}=(6.2\pm0.2)\times10^{42}~{\rm erg~s^{-1}}$, respectively. These two galaxies with bright [\ion{O}{3}] emissions, however, are undetected from the deep ALMA observations. 
We notice that Galaxy O3-004 is also detected in both H$\beta$ and Ly$\alpha$ ($z_{\rm Ly \alpha}=6.629$) from the NIRCam/WFSS observation and the MUSE observations \citep{Farina17}, respectively. The H$\beta$ and Ly$\alpha$ line luminosities are measured to be $L_{\rm H\beta}=(8.3\pm1.6)\times10^{41}~{\rm erg~s^{-1}}$ and $L_{\rm Ly\alpha}=(2.1\pm0.2)\times10^{42}~{\rm erg~s^{-1}}$, respectively. 
By assuming the standard Case B recombination with $T_e\sim10^4$ K and $n_e=10^2~{\rm cm^{-3}}$ and the absence of dust, we then estimate the Ly$\alpha$ escape fraction of this galaxy to be $f_{\rm esc}^{\rm Ly\alpha}=10\%$ which is similar with other high-redshift galaxies found in random field (i.e., a field without a luminous quasar) \citep[e.g.,][]{Hayes11}. 

These observations show that the environment of the earliest SMBHs is very complex with diverse galaxy populations and indicate that the early evolution of galaxies around the quasar is not simultaneous. This could also partially explain why the existing observations solely based on rest-frame UV emissions or ALMA observations could not recover galaxy overdensities around a large number of quasars \citep[see][for a review]{Overzier16}. 

\subsection{Implications to the formation and the environment of the earliest SMBHs} \label{sec:implication}
One of the key open questions in cosmology is how the billion-solar-mass BHs formed within just several hundred million years after the Big Bang. 
Growing these SMBHs requires a very massive seed BH growing continuously at the Eddington limit
and/or super-critical accretion on somewhat lighter seeds \citep[e.g.,][]{Volonteri12,Yang21}.
Theoretical models generally predict that the earliest billion-solar-mass BHs grow from remnants of quasi/supermassive stars \citep[generically called direct collapse BHs, DCBH; e.g.,][]{Begelman06,Inayoshi20,Sassano21} through cold flow accretion \citep[e.g.,][]{DiMatteo12} and sometimes also aided by mergers \citep[e.g.,][]{Li07, Dubois12, DiMatteo12}. 
This paradigm generally requires the seed to form in an overdense region, although not necessarily in the most massive halo of the region. This is because enhanced Lyman-Werner radiation from nearby sources \citep[e.g.,][]{Regan17} or dynamical heating \citep[e.g.,][]{Wise19}
are required to prevent gas fragmentation when it first collapses and steady gas flows are needed to sustain a rapid accretion of gas to grow the seed BH in during later stages \citep[e.g.,][]{Latif22}.

However, previous observations only confirmed a few galaxy overdensities around the earliest SMBHs from extensive ground-based and HST observations \citep[see][for a review]{Overzier16}, which seems inconsistent with the theoretical expectations. In order to explain the lack of galaxy overdensity around quasars observed in these studies, several works proposed that strong radiation from quasar could heat the surrounding gas and suppress the formation of galaxies in quasar vicinity \citep[e.g.,][]{Utsumi10} and/or the bulk of the galaxies in quasar vicinity are dusty \citep[e.g.,][]{Mazzucchelli17}. 
In this work, we found that J0305--3150, hosting a $\sim10^{9}~M_\odot$ SMBH, accreting at the Eddington limit and inhabiting a large-scale galaxy overdensity with both optical bright and dusty galaxies, is consistent with theoretical expectations. 
Earlier, shallower ground-based observations by \cite{Ota18} did not find a LAE galaxy overdensity in the J0305--3150 field. These observations indicate that lack of sensitivity, rather than radiation suppressing the formation of galaxies in quasar vicinity, is responsible for missing galaxy overdensities in previous shallower observations
\citep[][see also discussions in \S \ref{sec:structure}]{Habouzit19}, at least for the case of J0305--3150.
Therefore, we argue that it is still too early to judge whether the earliest SMBHs are in general good tracers of large-scale galaxy overdensities and whether quasar radiation could surpress galaxy formation in the quasar vicinity based on existing observations.  
The ASPIRE program, spectroscopically surveying 25 quasars in the EoR with excellent sensitivity and multi-wavelength coverage, will enable us to eventually resolve these questions.

\section{Summary} \label{sec:summary}
In this work, we provide a brief overview of the JWST ASPIRE program which will perform a legacy galaxy redshift survey in the fields of 25 reionization-era quasars using NIRCam/WFSS. From the early JWST observation of the field around the quasar J0305--3150, we discovered a filamentary structure traced by the quasar and 10 [\ion{O}{3}] emitters at $z=6.6$. This structure has a galaxy overdensity of $\delta_{\rm gal}=12.6^{+5.9}_{-5.0}$ over a 637 cMpc$^3$ volume, making it one of the most overdense structures found in the early Universe. 
By comparing with cosmological simulations, we argue that this filamentary structure traces an early overdensity, which could eventually evolve into a massive galaxy cluster. We also found that the most massive SMBHs in cosmological simulations generally trace galaxy overdensities but with a large variance on the galaxy numbers. This suggests that deep observations of a large sample of quasars (e.g., ASPIRE program) would be essential for a comprehensive understanding of the cosmic enviornment of the ealiest SMBHs.
Together with archival multi-wavelength observations, our JWST data indicates that the immediate vicinity of the quasar is very complex with a diverse galaxy population, including both UV bright galaxies and dusty star-forming galaxies. These observations highlight the needs of a multi-wavelength characterizations of galaxies around the earliest SMBHs. 
In the field of J0305--3150, we also discovered 31 [\ion{O}{3}] emitters at other redshifts, $5.3<z<6.7$, with half of them resides in two galaxy overdensities at $z=5.4$ and $z=6.2$. The serendipitous discovery of two galaxy overdensities in just one ASPIRE field indicates that high-redshift [\ion{O}{3}] emitters is a strongly clustered population.

\acknowledgments
FW thanks the support provided by NASA through the NASA Hubble Fellowship grant $\#$HST-HF2-51448.001-A awarded by the Space Telescope Science Institute, which is operated by the Association of Universities for Research in Astronomy, Incorporated, under NASA contract NAS5-26555. 
F.S. and E.E. acknowledge funding from JWST/NIRCam contract to the University of Arizona, NAS5-02105. 
JTS acknowledges funding from the European Research Council (ERC) Advanced Grant program under the European Union's Horizon 2020 research and innovation programme (Grant agreement No. 885301).
LB acknowledges support from NSF award AST-1909933 and NASA award \#80NSSC22K0808.
SEIB acknowledges funding from the European Research Council (ERC) under the European Union's Horizon 2020 research and innovation programme (grant agreement no. 740246 ``Cosmic Gas''.
L.C. acknowledges support by grant PIB2021-127718NB-I00,  from the Spanish Ministry of Science and Innovation/State Agency of Research MCIN/AEI/10.13039/501100011033
M.H. acknowledges support from the Zentrum f\"ur Astronomie der Universit\"at Heidelberg under the Gliese Fellowship.
Z.H. acknowledges support from US NSF grant AST-2006176.
H.D.J. was supported by the National Research Foundation of Korea (NRF) funded by the Ministry of Science and ICT (MSIT) of Korea (No. 2020R1A2C3011091, 2021M3F7A1084525, 2022R1C1C2013543).
GK is partly supported by the Department of Atomic Energy (Government of India) research project with Project Identification Number RTI~4002, and by the Max Planck Society through a Max Planck Partner Group.
AL acknowledges funding from MIUR under the grant PRIN 2017-MB8AEZ. 
S.R.R. acknowledges financial support from the International Max Planck Research School for Astronomy and Cosmic Physics at the University of Heidelberg (IMPRS–HD).
B.T. acknowledges support from the European Research Council (ERC) under the European Union's Horizon 2020 research and innovation program (grant agreement 950533) and from the Israel Science Foundation (grant 1849/19).
MT acknowledges support from the NWO grant 0.16.VIDI.189.162 (``ODIN'').

This work is based on observations made with the NASA/ESA/CSA James Webb Space Telescope. The data were obtained from the Mikulski Archive for Space Telescopes at the Space Telescope Science Institute, which is operated by the Association of Universities for Research in Astronomy, Inc., under NASA contract NAS 5-03127 for JWST. These observations are associated with program \#2078. Support for program \#2078 was provided by NASA through a grant from the Space Telescope Science Institute, which is operated by the Association of Universities for Research in Astronomy, Inc., under NASA contract NAS 5-03127.

This research is based in part on observations made with the NASA/ESA Hubble Space Telescope obtained from the Space Telescope Science Institute, which is operated by the Association of Universities for Research in Astronomy, Inc., under NASA contract NAS 5–26555. These observations are associated with program GO 15064.

This paper makes use of the following ALMA data: ADS/JAO.ALMA\#2017.1.01532.S. ALMA is a partnership of ESO (representing its member states), NSF (USA) and NINS (Japan), together with NRC (Canada), and NSC and ASIAA (Taiwan), in cooperation with the Republic of Chile. The Joint ALMA Observatory is operated by ESO, AUI/NRAO, and NAOJ.

This research made use of Photutils, an Astropy package for
detection and photometry of astronomical sources \citep{photutils}.
Some of the data presented in this paper were obtained from the Mikulski Archive for Space Telescopes (MAST) at the Space Telescope Science Institute. The specific observations analyzed can be accessed via \dataset[10.17909/vt74-kd84]{https://doi.org/10.17909/vt74-kd84}.

\vspace{5mm}
\facilities{ALMA, HST (ACS), HST (WFC3), JWST (NIRCam)}


\software{astropy \citep{Astropy},  
Matplotlib \citep{Matplotlib},
Numpy \citep{Numpy},
Photutils \citep{photutils},
Scipy \citep{Scipy},
Source Extractor \citep{SExtractor}
}



\addtocounter{figure}{-3}
\begin{figure*}[tbh]
\centering
\includegraphics[trim=0 0 0 0, clip, width=1\linewidth]{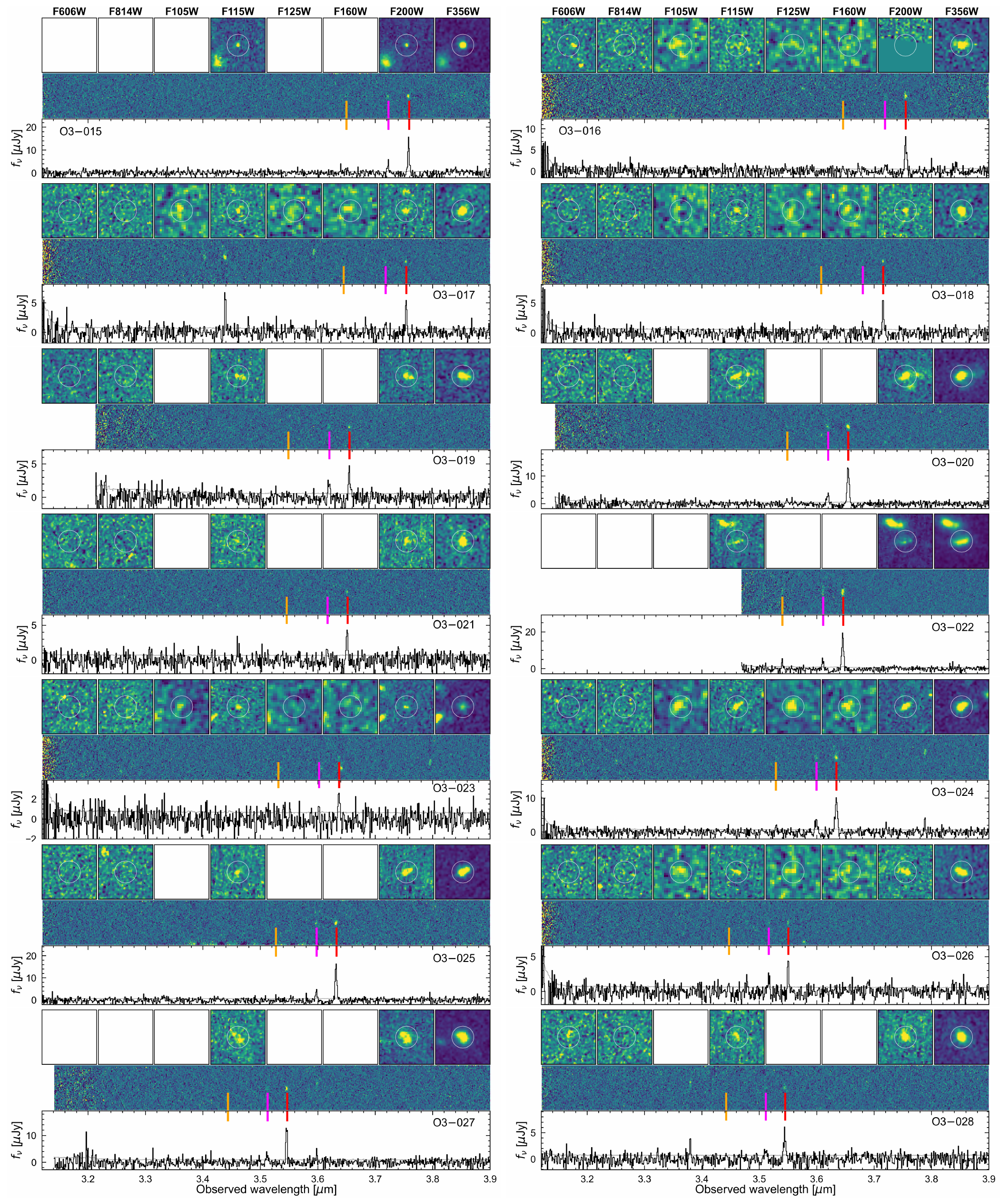}
\caption{ 
\textbf{Continued.}
}
\label{fig:spec2}
\end{figure*}

\addtocounter{figure}{-1}
\begin{figure*}[tbh]
\centering
\includegraphics[trim=0 0 0 0, clip, width=1\linewidth]{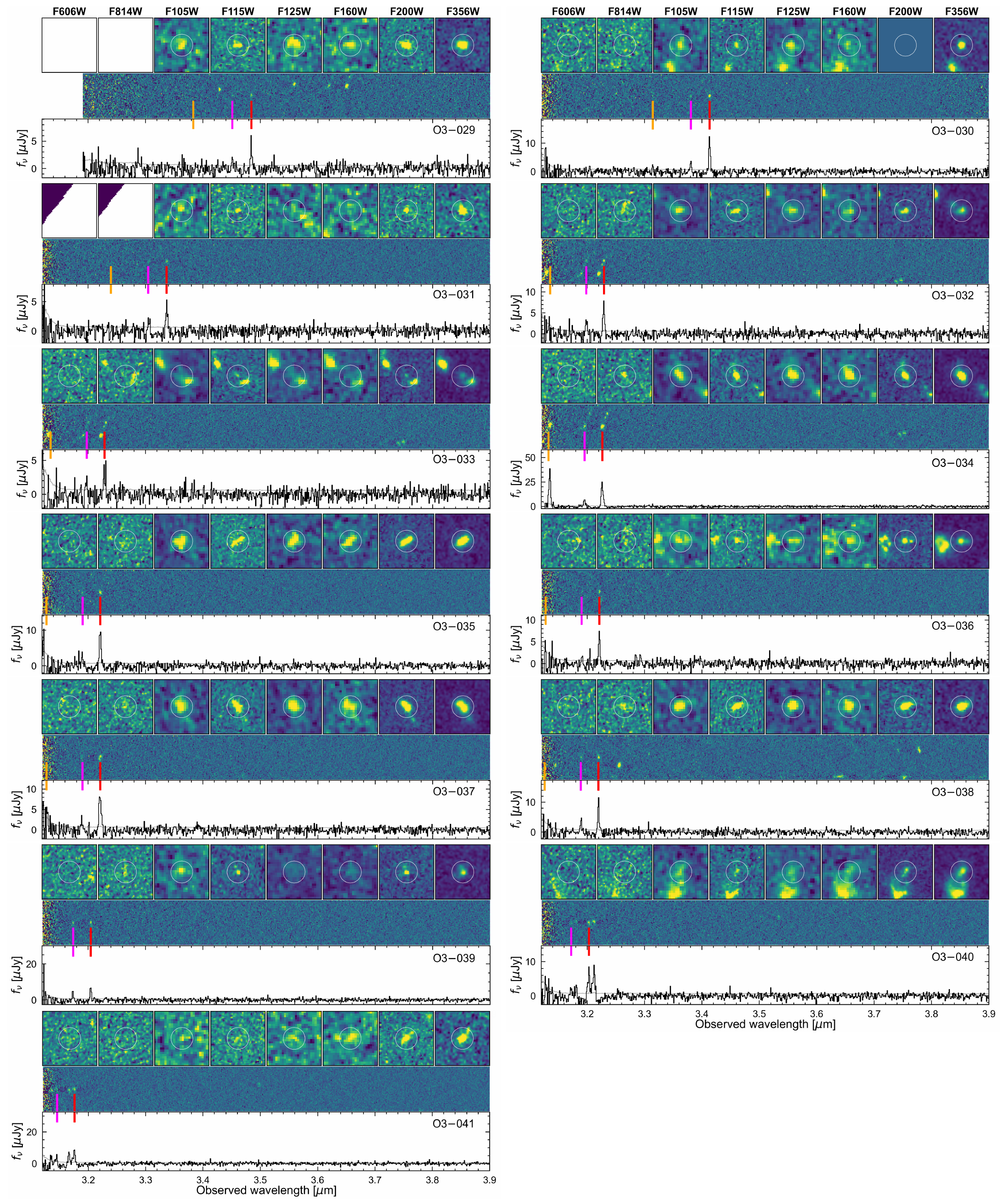}
\caption{ 
\textbf{Continued.}
}
\label{fig:spec3}
\end{figure*}


\bigskip




\end{document}